\documentclass[12pt]{article}
\usepackage{amssymb}

\begin{document}
\def\be{\begin{equation}}
\def\ee{\end{equation}}
\newcommand{\ds}{\displaystyle}
\newcommand{\Rea}{\mathbb{R}}
\newcommand{\Nat}{\mathbb{N}}
\newcommand{\Int}{\mathbb{Z}}
\newcommand{\Com}{\mathbb{C}}

\begin{titlepage}

\begin{flushright} CERN-TH/2001-238 \\
quant-ph/0109013 \end{flushright} \vspace{1.5cm}
\begin{center}
{\Large  How to Quantize  Phases and Moduli !} \vspace{0.5cm}
\\  {\large H.A.\
Kastrup\footnote{E-mail: Hans.Kastrup@cern.ch}} \vspace{0.4cm}
\\ {Theoretical Physics Division, CERN \\ CH-1211 Geneva 23,
Switzerland} \end{center}\vspace{0.4cm}

 \begin{center}{\bf Abstract}\end{center}
 A typical classical interference pattern of two waves with
 intensities $I_1,I_2$ and relative phase $\varphi =
 \varphi_2-\varphi_1$ may be characterized by the 3 observables
 $p=\sqrt{I_1\,I_2},\; p\cos\varphi,$ and $ -p\sin\varphi$. They are,
 e.g.\ the starting point for the semi-classical operational approach
  by Noh, Foug\`{e}res and Mandel (NFM) to
 the old and notorious phase problem in quantum optics.
  Following a recent group theoretical quantization of the symplectic
  space
  ${\cal{S}}= \{(\varphi \in \Rea \bmod{2\pi},p\,>\,0)\}$
 in terms of   irreducible unitary representations of the group
 $SO^{\uparrow}(1,2)$
the present paper applies those results to that controversial
problem of quantizing moduli and phases of complex numbers: The
Poisson brackets of the classical observables $p\cos
\varphi,-p\sin\varphi$ and $p>0$ form  the Lie algebra of the
group $SO^{\uparrow}(1,2)$.   The corresponding self-adjoint
generators $\widehat{p\cos \varphi}=K_1,
-\widehat{p\sin\varphi}=K_2$ and $\hat{p}=K_3$ of that group may
be obtained from its irreducible unitary representations. For the
positive discrete series the modulus operator $K_3$ has the
spectrum $\{n+k,n=0,1,2,\ldots; k>0\}$. Self-adjoint operators
$\widehat{\cos\varphi}$ and $\widehat{\sin\varphi}$ can  be
defined as $(K_3^{-1}K_1+K_1 K_3^{-1})/2$ and
--$(K_3^{-1}K_2+K_2K_3^{-1})/2$ which have  the theoretically
desired properties for $k \geq 0.5$. The approach advocated here
solves, e.g.\ the modulus-phase quantization problem for the
harmonic oscillator and appears to provide a full quantum
theoretical basis for the NFM-formalism.
\\ \\ PACS numbers: 03.65.Fd, 42.50.-p, 42.50.Dv
\end{titlepage}

\section{Introduction}
The problem how to quantize the modulus and the phase of a wave as
some kind of canonically conjugate variables and relate them to
genuine self-adjoint operators in Hilbert space is a very old one
and is - according to the still ongoing controversial discussions
in the field of quantum optics -  not yet settled (see, e.g.\ the
reviews \cite{rev}). A solution of that theoretical problem
becomes more and more urgent, however, because the experiments in
quantum optics are increasingly more refined and allow to
differentiate between different theoretical schemes.

Perhaps the most successful one of the schemes proposed up to now
is the semi-classical operational approach by Noh, Foug\`{e}res
and Mandel (NFM) \cite{Noh1,Noh2,Noh3,Noh4} which starts from
well-known classical interference concepts and reinterprets them
in terms of quantized observables. The approach works well as long
as the properties of the quantum theoretical ground state do not
become important, i.e.\ as long as one stays in the semi-classical
regime. The crucial - idealized - elements of the NFM -- scheme
which are of interest here are the following: \\ Consider the sum
\be A=a_1\,e^{i\,\varphi_1}+a_2\,e^{i\,\varphi_2}\ee of two
complex numbers $A_j$, where the phases $\varphi_j$ are chosen
such that $a_j
> 0,~j=1,2$. The quantities $a_j$ and $\varphi_j$ may be functions
of other parameters, e.g.\ space or/and time variables etc.
depending on the concrete experimental situation.
 The absolute
square of $A$ has the form \begin{eqnarray} w_3(I_1,I_2,
\varphi)&=&|A|^2(I_1,I_2,\varphi =\varphi_2-\varphi_1)=I_1+I_2+2\,
\sqrt{I_1\,I_2}\, \cos\varphi~,\\ && I_j =(a_j)^2\,,
j=1,2~.\nonumber
\end{eqnarray} Phase shifting one of the two amplitudes $A_j$ by an
 appropriate device yields new intensities: \begin{eqnarray}
w_4(I_1,I_2,\varphi) &=& w_3(\varphi +\pi)=
I_1+I_2-2\sqrt{I_1\,I_2}\cos\varphi~,\\ w_5(I_1,I_2,\varphi)
&=&w_3(\varphi +\pi/2)= I_1+I_2-2\sqrt{I_1\,I_2}\sin\varphi~,\\
w_6(I_1,I_2,\varphi)& =& w_3(\varphi- \pi/2)=
I_1+I_2+2\sqrt{I_1\,I_2}\sin\varphi~.\end{eqnarray} The essential
quantities for a classical description of the interference pattern
are then
\begin{eqnarray} 4P_1&=&w_3-w_4=
4\,p\,\cos\varphi\,,~p=\sqrt{I_1\,I_2}~,\\ 4P_2&=& w_5-w_6
=-4\,p\,\sin\varphi~,\\ 4P_3 &=& 4\,p=
\sqrt{(w_4-w_3)^2+(w_6-w_5)^2}
>0~. \end{eqnarray} The ratios $P_1/P_3$ and $P_2/P_3$ yield
$\cos\varphi$ and $-\sin\varphi$. \\ In quantum optics the
classical intensities $I_j, j=1,2,\,w_a, a=3,4,5,6$ become
energies of a  mean number of photons and the $w_a$   are formally
replaced by expectation values of number operators and so quantum
theory comes into play in a semi-classical way. \\ This NFM-scheme
has been quite successful, but justified theoretical doubts remain
as to the applicability of the approach for small values of
$|w_3-w_4|$ and $|w_5-w_6|$ and as to the commutativity of
corresponding quantum operators, especially of those corresponding
to $\cos\varphi$ and $\sin\varphi$.

In an attempt to find an appropriate quantized version of the
above classical description of interferences let me start with the
following observation: Suppose that $p>0$. As any  function
$f(\varphi,p)$ periodic in $\varphi$ with period $2\pi$ can --
under certain mathematical conditions -- be expanded in a Fourier
series and as $\cos(n\varphi)$ and $\sin(n\varphi)$ can be
expressed by polynomials of order $n$ in $\cos\varphi$ and
$\sin\varphi$, the observables $P_j\,,j=1,2,3,$ defined in Eqs.\
(6)-(8) are indeed the basic ones for such functions. Let \be
\{f_1,f_2\}=
\partial_{\varphi}f_1\,\partial_p f_2-\partial_pf_1\,
\partial_{\varphi}f_2 \ee
 be the Poisson bracket for any
two smooth functions $f_i(\varphi,p),\ i=1,2$. Then we have the
closed algebra
  \be \{P_3,P_1\}= -P_2,~~\{P_3,P_2\}=P_1,~~
\{P_1,P_2\}=P_3~, \ee which is just  the real Lie algebra of  the
group $SO^{\uparrow}(1,2)$ (identity component of the proper
Lorentz group in 2+1 space-time dimensions) or of one of its
infinitely many covering groups, e.g. the double covering
$SU(1,1)$ which is isomorphic to the group $SL(2,\Rea)$ and the
symplectic group $Sp(1,\Rea)$. Quantizing the classical
observables $P_j$ then consists in replacing them by the
self-adjoint generators $K_j \sim P_j$ of appropriate irreducible
unitary representations of those groups.

 The appropriate theoretical
background for this approach is provided by the so-called ``group
theoretical quantization'' which generalizes the usual
quantization procedure to systems which cannot be dealt with in
the  naive manner where classical canonical variable pairs are
replaced by multiplication and differential operators,
respectively. This approach to quantizing a classical system is a
genuine extension of the conventional method which is included as
a special case (see the reviews \cite{is1}).

 As an application of that generalized quantization scheme in the
present case of interest the symplectic manifold \be {\cal{S}}=
\{(\varphi \in \Rea \bmod{2\pi},p\,>\,0)\}\ee (associated with the
local symplectic form $ d\varphi \wedge dp$) was quantized  in
terms of the group $SO^{\uparrow}(1,2)$ for the purpose of
quantizing Schwarzschild black holes \cite{bo1}. We were not aware
then that the same quantization had been performed previously by
R.\ Loll in a different context \cite{lo}.

In the meantime I realized that this quantization also sheds new
light on the old unsolved problem how to represent phase and
modulus as self-adjoint operators in a Hilbert space associated
with a corresponding physical system (see the preliminary note
\cite{ka1}). Let me briefly point out some of the essential formal
features of the approach:

 The crucial point is that the manifold (11) has the nontrivial
  topology $S^1
\times \Rea^+,~\Rea^+$: real numbers $>0$. Such a manifold cannot
be quantized in the usual naive way used for a phase space with
the trivial topology $\Rea^2$ by converting a classical canonical
pair $(q,p)$ of phase space variables into operators and their
Poisson bracket into a commutator. Here the group theoretical
quantization scheme \cite{is1} as a generalization of the
conventional one  helps: The group $SO^{\uparrow}(1,2)$ acts
symplectically, transitively, effectively and (globally)
Hamilton-like on the manifold (11) (which may also be
characterized by the ``forward light cone''
$P_3^2-P_1^2-P_2^2=0\,, P_3 >0$) and, therefore, its irreducible
representations (or those of its covering groups) can provide the
basic self-adjoint quantum observables and their Hilbert space of
states (see Ref.\ \cite{bo1} for more details): In the course of
the group theoretical quantization one finds that the three basic
classical observables $P_j,j=1,2,3,$ correspond to the three
self-adjoint Lie algebra generators $K_j$ of a positive discrete
series irreducible unitary representation of the group
$SO^{\uparrow}(1,2)$ or one of its infinitely many covering
groups. The generators $K_j$ obey the commutation relations \be
[K_3,K_1]= i K_2,~~ [K_3,K_2]=-iK_1,~~[K_1,K_2]=-iK_3 ~~. \ee Here
$K_3$ is the generator of the compact sub-group $SO(2)$. \\ If the
minus sign in front of $i\,K_3$ on the r.h.s.\ of the last
commutator in Eqs.\ (12) is replaced by a plus sign we obtain the
Lie algebra of the rotation group or its covering group $SU(2)$.
It is crucial for the following discussions that we are  dealing
with the non-compact group $SO^{\uparrow}(1,2)$ instead.

 It is essential to realize
that a group theoretical quantization does $not$ assume that the
generators of the basic Lie algebra  themselves may be expressed
by some conventional canonical variables like in the case of
angular momentum. This may be the case locally in special
examples, but in general it will not be possible, especially not
globally. For more details see the discussion below and the Refs.\
\cite{is1,bo1,bo2}. The paper is organized as follows:

In section 2 I collect the essential elements as to the
self-adjoint Lie algebra generators of the irreducible unitary
representations of the group $SO^{\uparrow}(1,2)$ and  of its
covering groups and discuss important matrix elements of the
``observables'' $K_j\,, j=1,2,3,$ in the number state basis
(eigenstates of $K_3$). Then the operators $\widehat{\cos\varphi}$
and $\widehat{\sin\varphi}$ are introduced and some of their main
matrix elements in the number state basis calculated, too.

 Section
3 makes use of  $SO^{\uparrow}(1,2)$ Lie algebra related coherent
states (with $K_-|k,z\rangle= z\,|k,z\rangle$), introduced by
Barut and Girardello \cite{ba1}. Properties of simple matrix
elements are discussed and it is shown that the self-adjoint
$\widehat{\cos\varphi}$ and $\widehat{\sin\varphi}$ operators have
the right support -- the closed interval $[-1,+1]$ -- provided the
number $k>0$ which characterizes an irreducible unitary
representation has the lower bound $k\geq 0.5$.

Section 4 gives a physical interpretation of the previous results
without referring to ``field'' variables, i.e.\ without the use of
``underlying'' creation and annihilation operators or
corresponding ``modes''. It is possible - at least theoretically -
to show that a recorded interference pattern -- like in the
NFM-approach -- can be described satisfactorily in terms of {\em
observable quantities} like the intensities $I_1,I_2,P_1,P_2,P_3$
and their quantum mechanical counterparts, here especially the
operators $K_j\,,j=1,2,3,$ and functions of them.

In section 5 I  discuss briefly cases in which the Lie algebra
operators $K_j$ can be expressed in terms of creation and
annihilation operators. The most interesting one is that in which
the $K_j$ are expressed non-linearly in terms of one creation and
one annihilation operator $a^+$ and $a$ acting in the Fock space
of the harmonic oscillator. The general scheme immediately gives
``decent'' $\widehat{\cos\varphi}$ and $\widehat{\sin\varphi}$
operators (i.e.\ self-adjoint and with the correct spectrum),
solving an old and long discussed quantum mechanical problem
\cite{rev}!

\section{``Observable'' operators and their matrix elements in
the number state
basis}

In order to calculate expectation values and fluctuations we have
to know the actions of the operators $K_i,i=1,2,3,$ on the Hilbert
spaces associated with the positive discrete series of the
irreducible unitary representations of $SO^{\uparrow}(1,2)$ (or
its covering groups). In the following I rely heavily on Ref.
\cite{bo1} where more (mathematical) details and  Refs.\ to the
corresponding literature can be found.

As the eigenfunctions of $K_3$ -- the generator of the compact
subgroup $SO(2)$ of $SO^{\uparrow}(1,2)$ -- form a complete basis
of the associated Hilbert spaces, it is convenient to use them as
a starting point. The operators \be K_+=K_1+iK_2~,~~K_-=K_1-iK_2~
\ee $$ [K_+,K_-]=-2K_3\,,~[K_3,K_{\pm}]=\pm K_{\pm}~, $$ act as
ladder operators. The positive discrete series is characterized by
the property that there exists a state $|k,0\rangle$ for which $
K_-|k,0\rangle =0~.$ The number $k
>0$ characterizes the representation: For a general normalized
 eigenstate
 $|k,n\rangle$ of $K_3$
we have
 \begin{eqnarray} K_3|k,n\rangle&=&(k+n)|k,n\rangle~,~n=0,1,\ldots,
 \\
 K_+|k,n\rangle&=&\omega_n\,[(2k+n)(n+1)]^{1/2}|k,n+1\rangle~,~~
 |\omega_n|=1~, \\
 K_-|k,n\rangle&=&\frac{1}{\omega_{n-1}}[(2k+n-1)n]^{1/2}
 |k,n-1\rangle~.\end{eqnarray}
 In irreducible unitary representations the
operator $K_-$ is the adjoint operator of $K_+:\;
(f_1,K_+f_2)=(K_-f_1,f_2)$. The phases $\omega_n$ serve to
guarantee this property. Their choice depends on the concrete
realization of the representations. In the examples discussed in
Ref. \cite{bo1} they have the values 1 or $i$. In the following I
assume $\omega_n$ to be independent of $n:\omega_n=\omega$.

The Casimir operator  \be Q=K_1^2+K_2^2-K_3^2
=K_+K_-+K_3(1-K_3)=K_-K_+ - K_3(1+K_3) \ee has the eigenvalues
$q=k(1-k)$. The allowed values of $k$ depend on the group: For
$SO^{\uparrow}(1,2)$ itself one has $k=1,2,\ldots$ and for the
double covering $SU(1,1)$ $k=1/2,1,3/2,\ldots $. For the universal
covering group $k$ may be any real number $>0$ (for details see
Ref.\cite{bo1}). The appropriate choice will depend on the physics
to be described. In any case, for a unitary representation the
number $k$ has to be non-vanishing and positive!

The relation (15) implies \be |k,n\rangle
=\omega^{-n}\left[\frac{\Gamma(2k)}{n!\,\Gamma(2k+n)}\right]^{1/2}
(K_+)^n|k,0\rangle~.
\ee

The expectation values of the self-adjoint operators $ K_1 =(K_+
+K_-)/2$ and $K_2=(K_+-K_-)/2i $ (which correspond to the
classical observables $p\cos\varphi$ and $-p\sin\varphi$) with
respect to the eigenstates $|k,n\rangle$ and the associated
fluctuations may be calculated with the help of the relations
(13)-(16) : \be \langle k,n|K_i|k,n\rangle=0~,~~i=1,2~.\ee The
corresponding fluctuations are \be (\Delta K_i)^2_{k,n} =  \langle
k,n|K_i^2|k,n\rangle = \frac{1}{2}(n^2
+2nk+k)=\frac{1}{2}[(n+k)^2+q]~,~ i=1,2~. \ee Because of
$[K_1,K_2]=-iK_3$ the general uncertainty relation \be \Delta
A\;\Delta B \geq \frac{1}{2}|\langle \, |[A,B]|\, \rangle | \ee
for self-adjoint operators $A$ and $B$ here takes the special form
\be (\Delta K_1)_{k,n}\; (\Delta K_2)_{k,n} = \frac{1}{2}(n^2+2k
n+k) \geq \frac{1}{2}|\langle k,n|K_3|k,n\rangle
|=\frac{1}{2}(n+k)~. \ee The equality sign holds  for the ground
state $|k,n=0\rangle$.
\\ The eqs.\ (20) imply further (see also Eq.\ (17)) that \be
 \langle k,n|K_1^2|k,n\rangle +
\langle k,n|K_2^2|k,n\rangle = (n+k)^2 + q = \langle
k,n|K_3^2|k,n\rangle + q~. \ee This means that for very large $n$
the correspondence principle,
$(p\cos\varphi)^2+(p\sin\varphi)^2=p^2$, is fulfilled! For $q=0$
(i.e.\ $k=1$) we even have $K_1^2+K_2^2=K_3^2$!

  Next I define the self-adjoint operators \cite{ka1} $\widehat{
  \cos\varphi}$ and
  $\widehat{\sin\varphi}$ as follows: \be \widehat{\cos\varphi}=
  \frac{1}{2}(K_3^{-1}K_1
  +K_1K_3^{-1})~,~~
\widehat{\sin\varphi}= -\frac{1}{2}(K_3^{-1}K_2
  +K_2K_3^{-1})~. \ee
  Notice that $K_3^{-1}$ is well-defined because $K_3$ is a positive
   definite operator
  for the positive discrete series. One has $ K_3^{-1}|k,n\rangle =
  |k,n\rangle/(n+k)
  $. We shall demonstrate  below that the self-adjoint operators
  (24) have the right spectrum -- the full interval $[-1,+1]$ --
  provided $k \geq 0.5$.

Using the commutation relations (12) we obtain for the operators
(24):
\begin{eqnarray} {[} K_3, \widehat{\cos\varphi}{]}&=& -i\,
\widehat{\sin\varphi}~, \\ {[} K_3, \widehat{\sin\varphi}{]}&=&
i\, \widehat{\cos\varphi}~.
\end{eqnarray} The validity of these relations has been considered
 important for the properties of number,  $\cos$ and $\sin$
operators \cite{rev}. They were  introduced by Louisell \cite{lou}
who was the first to recognize that one should use the operator
versions of $\cos\varphi $ and $\sin\varphi$ instead of $\varphi$
itself in order to get a consistent quantization.
\\ The operators $\widehat{\cos\varphi}$ and
$\widehat{\sin\varphi}$ defined in Eq.\ (24) do not commute! Using
the Jacobi identity $[[A,B],C]+[[B,C],A]+[[C,A],B]=0$ we obtain
from Eqs. (25) and (26) \be
[[\widehat{\cos\varphi},\widehat{\sin\varphi}],K_3]=0~. \ee This
means that in an irreducible representation the commutator
$[\widehat{\cos\varphi},\widehat{\sin\varphi}]$ is a function of
$K_3$ and the Casimir operator $Q$,  that is to say the commutator
is diagonal in the number basis $|k,n\rangle$. We shall use this
property below in order to determine the commutator. \\ Similar
arguments lead to the relation \be
[(\widehat{\cos\varphi}^2+\widehat{\sin\varphi}^2),K_3]=0 ~,\ee
which we shall use in order to determine
$\widehat{\cos\varphi}^2+\widehat{\sin\varphi}^2$ which does not
have the classical value 1!

  From the Eqs.\ (13)-(16)  we get the important relations \begin{eqnarray}
  \widehat{\cos\varphi}|k,n\rangle &=& \frac{\omega}{4} f^{(k)}_{n+1}
  |k,n+1\rangle
  +\frac{1}{4\omega}f^{(k)}_{n}|k,n-1\rangle\;, \\
   \widehat{\sin\varphi}|k,n\rangle &=& -\frac{\omega}{4i} f^{(k)}_{n+1}
   |k,n+1\rangle
  +\frac{1}{4i\omega}f^{(k)}_{n}|k,n-1\rangle\;, \\f^{(k)}_n&=&
  [n(2k+n-1)]^{1/2}\left(
  \frac{1}{k+n}+\frac{1}{k+n-1}\right),\,f^{(k)}_{n=0} =0\,.\end{eqnarray}
  They imply
 \be \langle k,n|\widehat{\cos\varphi}|k,n\rangle=0~,~~
  \langle k,n|\widehat{\sin\varphi}|k,n\rangle=0~ \ee and \be
 \langle k,n|(\widehat{\cos\varphi})^2|k,n\rangle= \langle k,n|\,
 (\widehat{\sin\varphi})^2|k,n\rangle=\frac{1}{16}[(f^{(k)}_{n+1})^2+
 (f^{(k)}_{n})^2]\,,
 \ee \be  \langle k,n|[\,\widehat{\cos\varphi},\widehat{\sin\varphi}\,]
 |k,n\rangle =
 \frac{i}{8}[(f^{(k)}_{n+1})^2-(f^{(k)}_{n})^2]~. \ee
 This gives for the number states $|k,n\rangle$ the uncertainty
 relation \be
 (\Delta\widehat{\cos\varphi})_{k,n}\,(\Delta\widehat{\sin\varphi})_{k,n}=
\frac{1}{16}[(f^{(k)}_{n+1})^2+(f^{(k)}_{n})^2]\geq
\frac{1}{16}[(f^{(k)}_{n+1})^2-(f^{(k)}_{n})^2]~, \ee where again
the equality sign holds for the ground state $|k,n=0\rangle$! for
which
  \be
 \langle k,0|(\widehat{\cos\varphi})^2|k,0\rangle= \langle k,0|
 (\widehat{\sin\varphi})^2|k,0\rangle = \frac{(2k+1)^2}{8\,k(k+1)^2}~. \ee
 For $k=1/2$ the r.h.s.\ of Eq.\ (36) takes the value $4/9$ and for
 $k=1$ one has $9/32$.
It also follows that an upper bound $\langle
k,0|(\widehat{\cos\varphi})^2|k,0\rangle \leq 1$ implies for $k$
the lower bound $k \geq k_1\equiv [(0.5+0.5\sqrt{23/27}\,)^{1/3}
+(0.5-0.5\sqrt{23/27}\,)^{1/3}-1]/2 =0.162\ldots$. A slightly
higher lower bound for allowed values of $k$ will be discussed in
the next section.

  For very
 large $n$ we have the (correct) correspondence principle limits
 \begin{eqnarray}
 \langle k,n|(\widehat{\cos\varphi})^2|k,n\rangle&=& \langle k,n|
 (\widehat{\sin\varphi})^2|k,n\rangle\\ && \asymp \frac{1}{2}[1+
 \frac{1}{4}\frac{1+4q}{n^2}+O(n^{-4})]~~ \mbox{for}~~
 n\rightarrow \infty~,\nonumber \end{eqnarray}

 The relations (33) and (34) may now be used in order to calculate the
 commutator and the sum of their squares for the operators
 $\widehat{\cos\varphi}$ and $\widehat{\sin\varphi}$: We saw
 already above that (for an irreducible representation)
 these quantities have to be diagonal in the
 number basis $|k,n\rangle$! We just have to rewrite the r.h.\
 sides of the Eqs.\ (33) and (34) in terms of the diagonal operators
 $K_3$ and $Q$. The results are \be
[\widehat{\cos\varphi},\widehat{\sin\varphi}]=\frac{1}{4i}\;
\frac{K_3^2-1+2Q(2K_3^2-1)}{K_3(K_3^2-1)^2}~\ee and \be
\widehat{\cos\varphi}^2+\widehat{\sin\varphi}^2
=\frac{1}{4}\left[\frac{4K_3^4-7K_3^2+3}{(K_3^2-1)^2}+Q\,\frac{
4K_3^4-3K_3^2+1}{K_3^2(K_3^2-1)^2}\right]\,.
\ee The r.h.\ sides of these two operator relations seem to have a
problem if $K_3$ has the eigenvalue 1, i.e.\ for $k=1, n=0$.
However, this is not so: if one sandwiches those r.h.\ sides
between $\langle k,n=0|$ and $|k,n=0\rangle$ then all the factors
$k-1$ in the nominator and the numerator cancel nicely and one
gets  results according to Eqs.\ (33) and (34)!

For very large $n$ we get from Eq.\ (38) the correspondence limit
\be \langle k,n|[\,\widehat{\cos\varphi},\widehat{\sin\varphi}\,]
|k,n\rangle =\frac{1}{4i}\frac{1+4q}{n^3}+ O(n^{-5})~~
\mbox{for}~~
 n\rightarrow \infty~. \ee
 I would like to point out that  phase operators
 associated with the Lie algebra of the group $SO^{\uparrow}(1,2)$
different from those above have been discussed by other authors
\cite{ge1,ag1,vo1,bri,bo2}.
\section{Matrix elements with respect to coherent states}
  Next I discuss some properties of coherent states. Contrary to the
  conventional
 coherent states (i.e.\ the eigenstates of the Bose annihilation operator
  associated
 with the harmonic oscillator, see e.g.\ the reviews \cite{kl1,pe1,zha}
  and the
  modern exposition
 \cite{ha1}) there are several inequivalent
 ways  \cite{pe1} \cite{tri} to define coherent states related to
  the group $SO^{\uparrow}(1,2)$ or $SU(1,1)$
(see also the Refs.\ \cite{su11}). For our purposes the
  definition  \be K_-|k,z\rangle
 =z\,|k,z\rangle~,~~ z=\rho\,e^{i\phi} \in \Com ~, ~ \rho=|z|\,.\ee seems to
  be an interesting one, at least theoretically.
  About the possibility of their experimental realization I have nothing to
  say! The states (41), introduced by Barut and Girardello
  \cite{ba1}, have widely been discussed in the literature
  concerned with quantum optical applications of the group $SU(1,1)$
   \cite{bri,squ} and most
  of the general results I shall mention in the following are
  well-known.

 Using the property (18) we get \be \langle k,n|k,z\rangle=
  \frac{1}{\bar{\omega}^n}
\left[\frac{\Gamma(2k)}{n!\,\Gamma(2k+n)}\right]^{1/2}z^n\,
\langle k,n=0|k,z\rangle~, \ee ($\bar{\omega}:~\mbox{compl.\
conj.\ of}~\omega$) so that \begin{eqnarray} \langle
k,z|k,z\rangle& =& \sum_{n=0}^{\infty}= \langle k,z|k,n\rangle
\langle k,n|k,z\rangle\\  &=& \Gamma(2k)|\langle
k,n=0|k,z\rangle|^2 \sum_{n=0}^{\infty}\frac{\rho^{2n}}{n!\,
\Gamma(2k+n)}\nonumber \\ &=&\Gamma(2k)|\langle
k,n=0|k,z\rangle|^2 \rho^{1-2k}\,I_{2k-1}(2\rho)~, \nonumber
\end{eqnarray} where \be I_{\nu}(x) = \left
(\frac{x}{2}\right)^{\nu}\sum_{n=0}^{\infty}
\frac{1}{n!\,\Gamma(\nu+n+1)}\left (\frac{x}{2}\right)^{2n} \ee is
the usual modified Bessel function of the first kind \cite{er1}
which has the asymptotic expansion \begin{eqnarray} I_{\nu}(x)
&\asymp& \frac{e^x}{\sqrt{2\pi\,x}}\left[1-\frac{4\nu^2 -1}{8\,x}+
2\frac{(4\nu^2-1)(4\nu^2-9)}{16^2\,x^2}+ O(x^{-3})\right]\\
&&~~~~~~~~~~~~ \mbox{for}~x\rightarrow +\infty ~. \nonumber
\end{eqnarray} If $\langle k,z|k,z\rangle=1$ we have \be|\langle
k,n=0|k,z\rangle|^2 \equiv |C_z|^2 =
\frac{\rho^{2k-1}}{\Gamma(2k)\, I_{2k-1}(2\rho)}~. \ee Choosing
the phase of $C_z$ appropriately and absorbing the phase $\omega$
into a redefinition of $z$ we finally get the expansion \be
|k,z\rangle=
\frac{\rho^{k-1/2}}{\sqrt{I_{2k-1}(2\rho)}}\sum_{n=0}^{\infty}
\frac{z^n}{\sqrt{n!\,\Gamma(2k+n)}}\,|k,n\rangle ~~.\ee Notice
that $|k,z=0 \rangle = |k,n=0\rangle$. \\  Two different coherent
states  are not orthogonal: \begin{eqnarray} \langle k, z_2|k,z_1
\rangle &=& \sum_{n=o}^{\infty}\langle k, z_2|k,n \rangle  \langle
k,n|k,z_1 \rangle \nonumber \\ &=& \frac{(\bar{z}_2z_1)^{1/2
-k}|z_2z_1|^{k-1/2}\,I_{2k-1}(2\sqrt{
\bar{z}_2z_1})}{\sqrt{I_{2k-1}(2|z_2|)I_{2k-1}(2|z_1|)}}~.
\end{eqnarray} They are complete, however, in the sense that, with
$z=\rho\,e^{i\,\phi}$, we have the  relation
\begin{eqnarray}&& \frac{2}{\pi}\int_0^{\infty}d\rho\, \rho\,
K_{2k-1}(2\rho)\,I_{2k-1}(2\rho) \int_0^{2\pi}d\phi\, \langle
k,n|k,\rho\, e^{i\phi}\rangle \langle k,  \rho\, e^{i\phi}|k,n
\rangle \nonumber \\ & & ~~~~~~~~~~~~~~~  =\langle k,n|k,n\rangle
=1\end{eqnarray} because \cite{GR} $$ \int_0^{\infty} d\rho\,
\rho^{2(n+k)}K_{2k-1}(2\rho)=\frac{1}{4}n!\,\Gamma(2k+n)~. $$ Here
$K_{\nu}(x)$ is the modified Bessel function of the third kind
\cite{er1}.

 The following expectation values are associated with the
states $|k,z\rangle$:
\begin{eqnarray} \langle K_3\rangle_{k,z} &\equiv& \langle
 k,z|K_3|k,z\rangle=
k+\rho\, b_k(\rho)\,, \\ && b_k(\rho)=
\frac{I_{2k}(2\rho)}{I_{2k-1}(2\rho)}~,
\\ \langle K_3^2\rangle_{k,z}&=& k^2+\rho^2+ \rho\,
b_k(\rho)~,\end{eqnarray} so that \be (\Delta
K_3)^2_{k,z}=\rho^2(1-b_k^2(\rho))+(1-2k) \rho\,b_k(\rho)~.\ee For
$\rho \to 0$ one has \be b_k(\rho) \to \rho~~\mbox{for}~\rho\to 0
\ee and for very large $\rho$ we get from Eq.\ (45) that \be
b_k(\rho)\asymp 1 +\frac{1-4k}{4\rho}+
\frac{3-16q}{32\rho^2}+O(\rho^{-3})\, \ee and therefore for the
r.h.\ sides of the relations (50) and (53) the leading terms \be
\langle K_3\rangle_{k,z} \asymp \rho +\frac{1}{4}+O(\rho^{-1})~,~~
(\Delta K_3)^2_{k,z} \asymp \frac{1}{2}\rho + O(\rho^{-1})
~\mbox{for}~\rho\rightarrow +\infty~~. \ee This, together with the
probability \be |\langle k,n|k,z \rangle|^2 =
\frac{\rho^{2(n+k)-1}}{n!\,\Gamma(2k+n)\,I_{2k-1}(2\rho)}\asymp 2
\sqrt{\pi} \frac{\rho^{2(n+k)-1/2}}{n!\,\Gamma(2k+n)}e^{-2\rho}
\ee shows that the corresponding distribution for large $\rho$ is
not Poisson-like!

In addition we have the following expectation values:
\begin{eqnarray} \langle K_1\rangle_{k,z}
&=&\frac{1}{2}(\bar{z}+z)=\rho\,\cos\phi\,,~ \langle
K_2\rangle_{k,z} =\frac{1}{2i}(\bar{z}-z)=-\rho\,\sin\phi~,\\
\langle K_1^2\rangle_{k,z}&=& \rho^2\cos^2\phi +\frac{1}{2}\langle
K_3\rangle_{k,z}\,, ~ \langle K_2^2\rangle_{k,z} =
\rho^2\sin^2\phi +\frac{1}{2}\langle K_3\rangle_{k,z}\,, \\ &&
\langle K_1^2+ K_2^2\rangle_{k,z} = \rho^2 + \langle
K_3\rangle_{k,z}\,, \nonumber
\\(\Delta K_1)^2_{k,z}& =&(\Delta K_2)^2_{k,z} =\frac{1}{2}\langle
K_3\rangle_{k,z}~~.\end{eqnarray} In deriving the relations (59)
the equality $K_-K_+= K_+K_-+2K_3$ (Eqs.\ (13)) has been used.

Comparing Eqs.\ (6),(7) and (58) we see the close relationship
between the expectation values $\langle
K_i\rangle_{k,z}\,,\,i=1,2,$ and their classical counterparts.
This supports the above choice (41) as coherent states. Further
support comes from their property to realize the minimal
uncertainty relation: From the third commutator in Eqs.\ (12) we
get the general inequality \be (\Delta K_1)^2 (\Delta K_2)^2 \geq
\frac{1}{4}|\langle K_3 \rangle|^2~. \ee The relations (60) show
that the coherent states (41) realize the minimum of the
uncertainty relation (61). One can, of course, extend the
discussion to associated squeezed states \cite{bri,squ}. Their use
in the present context will be of considerable interest.

For  matrix elements of the operators $\widehat{\cos\varphi}$ and
$\widehat{\sin\varphi}$ we get (from Eqs.\ (29), (30) and (47))
\begin{eqnarray}  \langle k,n|\widehat{\cos\varphi}|k,z\rangle& =&
\frac{\rho^{k-1/2}}{4\sqrt{I_{2k-1}(2\rho)}}\left[\frac{z^{n-1}
}{\sqrt{(n-1)!\Gamma(2k+n-1)}}f^{(k)}_n
 \right. \\ &&\left.
+\frac{z^{n+1}}{\sqrt{(n+1)!\Gamma(2k+n+1)}}f^{(k)}_{n+1}\right]~,
\nonumber \\ \langle k,n|\widehat{\sin\varphi}|k,z\rangle& =&
\frac{\rho^{k-1/2}}{4i\sqrt{I_{2k-1}(2\rho)}}\left[-\frac{z^{n-1}
}{\sqrt{(n-1)!\Gamma(2k+n-1)}}f^{(k)}_n \right.  \\ && \left. +
\frac{z^{n+1}}{\sqrt{(n+1)!\Gamma(2k+n+1)}}f^{(k)}_{n+1}\right]~.
\nonumber \end{eqnarray} From them and (47) we obtain
 the expectation values
\begin{eqnarray} \langle \widehat{\cos\varphi}\rangle_{k,z} &=&
\frac{\bar{z}+z}{2\rho}\,\frac{g^{(k)}(\rho)}{I_{2k-1}(2\rho)}=
\cos\phi \,\frac{g^{(k)}(\rho)}{I_{2k-1}(2\rho)} ~,\\ \langle
\widehat{\sin\varphi}\rangle_{k,z}&=&
\frac{-\bar{z}+z}{2i\rho}\,\frac{g^{(k)}(\rho)}{I_{2k-1}(2\rho)}=
\sin \phi\, \frac{g^{(k)}(\rho)}{I_{2k-1}(2\rho)}~,\\&&
g^{(k)}(\rho)= \frac{1}{2}
\sum_{n=0}^{\infty}\frac{\rho^{2(n+k)}}{n!\,\Gamma(2k+n)}\left(\frac{1}{n+k}
+\frac{1}{n+k+1}\right)\,. \end{eqnarray} One has \be
g^{(k)}(\rho)= \frac{1}{2} \int_0^{2\rho}du\,
I_{2k-1}(u)+\frac{1}{8\rho^2}\int_0^{2\rho}du\, u^2I_{2k-1}(u)~.
\ee The right hand side may be expressed by a combination of
modified Bessel and Lommel functions \cite{lu1}. For large $\rho$
one obtains \cite{lu2} \be \frac{g^{(k)}(\rho)}{I_{2k-1}(2\rho)}
\asymp 1-\frac{1}{4\rho} + O(\rho^{-2})~ \mbox{for large}~\rho~
\ee which again gives the expected correspondence principle limits
for $\langle \widehat{\cos\varphi}\rangle_{k,z}$ and $\langle
\widehat{\sin\varphi}\rangle_{k,z}$. Notice that the ratio
$\langle \widehat{\sin\varphi}\rangle_{k,z}/\langle
\widehat{\cos\varphi}\rangle_{k,z}= \tan \phi$ is independent of
$k$ and $\rho$!

 Other expectation values like
$\langle \widehat{\cos\varphi}^2\rangle_{k,z}$ etc.\ may be
calculated by observing that $$\langle
\widehat{\cos\varphi}^2\rangle_{k,z}= \sum_n \langle
k,z|\widehat{\cos\varphi}|k,n\rangle \langle k,n|
\widehat{\cos\varphi}|k,z\rangle= \sum_n | \langle k,n|
\widehat{\cos\varphi}|k,z \rangle|^2   $$ etc. The resulting
expressions are not simple.

 The operators $\widehat{\cos
\varphi}$ and $\widehat{\sin \varphi}$ are bounded self-adjoint
operators (see their definition (24) and Eqs.\ (33)). They have a
 continuous spectrum covering the interval $[-1,+1]$ for $k \geq
0.5$. The last assertion follows from Eqs.\ (64) and (65) together
with a numerical analysis of the ratio
$g^{(k)}(\rho)/I_{2k-1}(2\rho)$ which shows that ratio  to be $<
1$ for all finite $\rho$ if $k \geq 0.5$. That is not so, e.g.\
for $k=0.25$ for which  $g^{(k)}(\rho)/I_{2k-1}(2\rho)$ becomes
larger than 1 for certain $\rho$-values. The bound $k \geq 0.5$ is
definitely a sufficient one because $g^{(k)}(\rho)/I_{2k-1}(2\rho)
\leq 1$ numerically also for $k=0.4$. These - not yet very
detailed - numerical results and the relation (68) imply that at
least for $k \geq 0.5$ we have \be \sup_z|\langle
\widehat{\cos\varphi}\rangle_{k,z}|= \sup_z |\langle
\widehat{\sin\varphi}\rangle_{k,z}|=1~~ \mbox{for}~k \geq 0.5 \ee
from which the support of the spectrum follows \cite{ree}. Thus,
for the groups $SO^{\uparrow}(1,2)$ and $SU(1,1)$ which have $k=1$
and $k=1/2$ respectively as their lowest $k$-values we are on the
safe side.\\ The reason for a lower bound for $k$ in order to
ensure that the operators $\widehat{\cos\varphi}$ and
$\widehat{\sin\varphi}$ have the right spectrum can be seen
qualitatively from the first ($n=0$) term in the series (66). That
term (divided by $I_{2k-1}$) diverges for $k \to 0$.

 The ansatz $ |\mu\rangle =
\sum_{n=0}^{\infty}a_n\,|k,n\rangle $ for the improper
``eigenfunctions'' of
 $\widehat{\cos \varphi}$ with ``eigenvalues'' $\mu, ~
 \widehat{\cos \varphi}|\mu\rangle = \mu\,|\mu \rangle~,$ leads to the
  recursion
 formula $ a_{n+1}=(4\,\mu\,a_n
 -f^{(k)}_{n}\,a_{n-1})/f^{(k)}_{n+1}\,,~f^{(k)}_{0}=0\,,$ which allows to
  express the $a_n$ by
 $\mu, a_0$ and the $f^{(k)}_n$. \section{Interpretation in terms
 of ``observables''}
 Let me start with a very important remark which
 is crucial for the following physical interpretation of the results in the
 last two sections:

 In the case of a {\em group theoretical quantization} of a
 classical system it is $not$ required that the generators of the
 corresponding Lie algebra are expressible in terms of pairs
 $(\hat{q}_j,\hat{p}_j)\,, j=1,2,\ldots$ of canonical operators or
  in terms of the associated
annihilation and creation operators $a_j =
(\hat{q}_j+i\,\hat{p}_j)/2\,, a_j^+ = (\hat{q}_j-i\,\hat{p}_j)/2$.
A well-known example is the angular momentum: Its components
$\hat{l}_j\,,j=1,2,3,$ are expressible in terms of the 3 pairs
$(\hat{q}_j,\hat{p}_j)$ but this is not essential at all for the
quantum theory of the angular momentum. That can be constructed
from the single property that the 3 operators $\hat{l}_j$ generate
the Lie algebra of the group $SO(3)$ or that of its covering group
$SU(2)$ the representations of which allow for half-integer spins
not expected from semi-classical arguments ! Classically the
Poisson brackets of the 3 components $l_1 = q_2p_3-q_3p_2\,,l_2
=\ldots$ fulfill the $SO(3)$ Lie algebra, too. However, this
applies to the orbital angular momentum only.

Let us apply a similar analysis to the quantities (1)-(8) from the
introduction: On the classical level the actual ``observables''
for a complete description of the interference pattern are the two
intensities $I_1\,, I_2$ and the 3 quantities $P_j\,,j=1,2,3$. The
latter are not independent because
\be
P_1^2+P_2^2=P_3^2\,,~\cos\varphi=P_1/P_3\,,~\sin\varphi=-P_2/P_3\,.
\ee The individual phases $\varphi_j\,, j=1,2,$ do not have to be
known, only their difference $\varphi=\varphi_2-\varphi_1$. All
those observables may be determined by measuring the 6 intensities
$I_1,I_2$ and $w_a\,, a=3,4,5,6$. $I_1$ and $I_2$ may be measured
by shielding one of the two interfering waves completely. Notice
that $P_3^2=p^2 $ may also be obtained from $(I_1+I_2)^2$: \be
2p^2=(I_1+I_2)^2-I_1^2-I_2^2\,,~I_1+I_2= w_3+w_4=w_5+w_6\,. \ee We
are, of course, assuming a very idealized situation, namely that
all the quantities are stationary in time, that there is no
absorption in the $\lambda/2$ and $\lambda/4$ phase shifters etc.\
etc.

As a mere pedagogical example take Young's interference experiment
with a completely coherent source the light of which has a
definite frequency and falls onto two (equal) pinholes as origins
for the two waves which interfere on a screen on the other side of
the pinholes where the interference pattern is recorded
\cite{you}. At a certain position on the screen one has the
intensity $w_3$. When shielding pinhole 1 one observes the
intensity $w_3=I_2$ and when shielding pinhole 2 one observes
$I_1$. Placing an ideal $\lambda/2$ phase shifter behind pinhole 1
or 2 yields $w_4$. Using instead a $\lambda/4$ shifter gives $w_5$
and placing in addition a $\lambda/2$ shifter in behind the other
hole yields $w_6$.

Let us turn now to the quantum theory of the system \cite{wa}: \\
Crucial new features are that the 3 self-adjoint operators
$K_j=\hat{P}_j$ do no longer commute but obey the Lie algebra (12)
and that $K_3$ which corresponds to the classical quantity
$\sqrt{I_1\,I_2}$ is not a pure number operator! $K_3$ has the
eigenvalues $n+k\,,n=0,1,\ldots$ and the value of $k>0$ depends on
the type of global group one associates with the Lie algebra (12).
As already mentioned before: for the group $SO^{\uparrow}(1,2)$
one has $k=1$, for $SU(1,1)~ k=1/2$ etc.\ (for more details see
Ref.\ \cite{bo1}). If we define the number operator \be N=K_3-k\,
\mbox{I}\,,~ \mbox{I}:~\mbox{unity operator}\,, \ee then we have
the commutators \be [N,K_1]=i\,K_2\,,~[N,K_2]=-i\,K_1\,,~
[K_1,K_2]=-i\,(N+k\cdot\mbox{I})\,, ~k>0\,, \ee which is a Lie
algebra formed by $K_1\,,K_2$ and $N$ with a central extension
\cite{go} characterized by $k$. In the following the unity
operator I is no longer exhibited explicitly. \\ Instead of the
classical relation (70) we now have from Eq.\ (17) for an
irreducible representation \be K_1^2+K_2^2=K_3^2 +q =N^2+(2N+1)\,k
\ee which requires a corresponding modification of the NFM-type
analysis for small numbers $n$ of the quanta where the
non-vanishing $k$ makes itself felt! \\ In a quantum optical
experiment where mean photon $numbers$ $\bar{n}_a
=\langle\,|N_a|\,\rangle,$ $ a=3,4,5,6,$ are recorded instead of
the intensities $w_a$ one expects the relations (6) and (7) to
have the quantum expectation value correspondences \be
\bar{n}_3-\bar{n}_4= 4\,\langle \, |K_1|\, \rangle \,, \ee and \be
\bar{n}_5-\bar{n}_6= 4\,\langle \, |K_2|\, \rangle \,, \ee where
$|\,\rangle$ is a general state and the $N_a$ some appropriate
number operators corresponding to the quantum version of the
$w_a$. \\ However, because of Eq.\ (74) one can no longer expect a
simple correspondence relation for the classical equality (8). For
a state $|\,\rangle$ in an irreducible representation on has from
Eq.\ (74) \be \langle\, |K_1^2+K_2^2|\, \rangle  = \langle \,
|N^2|\, \rangle  + k\,(2\langle \, |N|\, \rangle  +1)\,.\ee On the
other hand one expects
\begin{eqnarray} \langle \, |K_1^2|\, \rangle
&=&\frac{1}{16}\langle \, |(N_3-N_4)^2|\, \rangle \,,\\ \langle \,
|K_2^2|\, \rangle  &=&\frac{1}{16}\langle \, |(N_5-N_6)^2|\,
\rangle \,,
\end{eqnarray} In addition one might get information about
$\langle \, | K_3^2|\, \rangle $ from the quantum version of Eq.\
(71): \begin{eqnarray} 2\, \langle \, |K_3^2|\, \rangle & =&
\langle\,| (N_3+N_4)^2|\, \rangle  -\langle \, |N_1^2|\, \rangle
-\langle \, |N_2^2|\, \rangle \,, \\ 2\, \langle \, |K_3^2|\,
\rangle & =& \langle\,| (N_5+N_6)^2|\, \rangle  -\langle \,
|N_1^2|\, \rangle -\langle \, |N_2^2|\, \rangle \,. \nonumber
 \end{eqnarray} As an additional general
information one has the inequality \be \Delta K_1\cdot \Delta K_2
\geq \frac{1}{2}\langle \, |K_3|\, \rangle \,. \ee In this
elementary approach there appears to be no obvious way to measure
$\langle \, |K_3=N+k|\, \rangle $ itself directly for an arbitrary
state $|\, \rangle $.

 Let us see next how
these more general considerations look for the special states
$|k,n\rangle$ and $|k,z\rangle$ from sections 2 and 3. I have
nothing to say about their possible experimental realizations!

In case of the number eigenstates $|k,n\rangle $ we have the
relations (19) for the expectation values of $K_1$ and $K_2$ and
(32) for those of $\widehat{\cos\varphi}$ and
$\widehat{\sin\varphi}$, i.e.\ there is no interference pattern at
all! According to Eqs.\ (75) and (76) this should correspond to
the relations $\bar{n}_4=\bar{n}_3\,,\bar{n}_6=\bar{n}_5$ The
associated fluctuations for $K_j\,,j=1,2,$ are given by Eq.\ (20)
and their sum by Eq.\ (23) which relates those fluctuations to the
exact eigenvalues $n+k$ of $K_3$ and $(n+k)^2$ of $K_3^2$.
Supplemented with the relations (80) this should allow for a
determination of $n$ and, very important, of $k$. The value of the
non-vanishing positive parameter $k$ which characterizes the
properties of the ground state -- see e.g.\ Eq.\ (36) -- does not
appear to be determined by general considerations alone, at least
not to me. So it should be determined or confirmed experimentally,
like in the case of  the ground state for the harmonic oscillator.
\\ If one knows $n$ and $k$ then one can infer the fluctuations of
$\widehat{\cos\varphi}$ and $\widehat{\sin\varphi}$ from Eq.\
(33).

In the case of the coherent states $|k,z\rangle$ the situation is
quite pleasant -- theoretically: With the help of Eqs.\ (75), (76)
and (58) one determines $\rho\,\cos\varphi$ and $\rho
\sin\varphi$. $\rho$ itself can in addition be obtained by
combining the relations (78), (79) and (59). These determine
$\langle K_3\rangle_{k,z}$ and, because of Eq.\ (74), $\langle
K_3^2\rangle_{k,z}$ as well. The results may be cross-checked by
means of the relations (80)!

 It appears that the emerging picture of
describing the quantum theory of moduli and phases in interference
experiments in terms of the Lie algebra of the group
$SO^{\uparrow}(1,2)$ or one of its (infinitely many) covering
groups is quite promising and may lead to progress in that field!

 Up to now I have not specified the concrete form of the Hilbert space, the
 operators $K_i,i=1,2,3,$ and the eigenfunctions $|k,n\rangle$ and
 $|k,z\rangle$. Several
 interesting examples may be found in Ref.\
 \cite{bo1} and the coherent states $|k,z\rangle$ can be constructed
 explicitly from the concrete
 form of the operators $K_-$ given there. Many examples are also
 contained in  numerous of the Refs. \cite{su11}.
\section{Interpretation in terms of creation and annihilation
operators: phase operators for the harmonic oscillator}
 In the following I discuss three -- well-known -- examples in which
 the 3 Lie algebra operators $K_j$ are expressed in terms of the
 ``beloved'' bosonic creation and annihilation operators $a^+$ and
 $a$\, with \be [a\,, a^+]=1\ee which act on a n-quanta state
 $|n\rangle $ as  \be a^+|n\rangle =
 \sqrt{n+1}\,|n+1\rangle\,~a\,|n\rangle
 =\sqrt{n}\,|n-1\rangle\,,~N\,|n\rangle=n\,|n\rangle,\,~N=a^+a\,. \ee
 Comparing
 these relation with the Eqs.\ (14)-(16) suggests the
 ansatz  \be K_+ = a^+\sqrt{N+2k}\,,~K_-=
 \sqrt{N+2k}\,a\,,~K_3=N+k\,. \ee Using the commutation
 relation (82) it is easy to verify that the operators (84) have the
 properties (14)-(16) and form the Lie algebra (13). We know from
 our general discussion that the operators (84) are self-adjoint
 in the Fock space of the harmonic oscillator
 provided $k \geq 0.5$. The operators $\widehat{\cos\varphi}$ and
 $\widehat{\sin\varphi}$ here take the explicit forms
\begin{eqnarray}\widehat{\cos\varphi}&=&\frac{1}{4}\frac{1}{N+k}
[8a^+\sqrt{N+2k}+
 \sqrt{N+2k}\,a]+\\ &&+\frac{1}{4}[a^+\sqrt{N+2k}+
 \sqrt{N+2k}\,a]\frac{1}{N+k}\,, \nonumber \\
 \widehat{\sin\varphi}&=&\frac{i}{4}\frac{1}{N+k}[a^+\sqrt{N+2k}-
 \sqrt{N+2k}\,a]+\\&&+\frac{i}{4}[a^+\sqrt{N+2k}-
 \sqrt{N+2k}\,a]\frac{1}{N+k}\,.\nonumber \end{eqnarray}
Again, according to our general results these operators are
self-adjoint with a spectrum in the interval $[-1,+1]$ provided
$k\geq 0.5$ (as mentioned above: this is a sufficient lower bound
which was found by numerical methods. For $k=0.25$ the same
analysis shows that the spectrum exceeds that interval!) For
$k=1/2$ or $k=1$ we are on the safe side and therefore the
operators are decent self-adjoint $\cos\varphi$ and $\sin\varphi$
operators for the harmonic oscillator! Actually there is no
obvious reason up to now to identify the modulus operator $K_3$
with the Hamiltonian $H=N+1/2$ of the harmonic oscillator though
one might be inclined to do so. The relations (25) and (26) here
follow immediately from $[N\,,a^+]=a^+$ and $[N\,,a]=-a$. \\ If we
compare the expressions (85) and (86), e.g.\ for $k=1/2$ or $k=1$,
with the questionable operators previously suggested by Dirac (and
Heitler), \be \widehat{\cos\varphi}_D = \frac{1}{2}(a\,N^{-1/2}+
N^{-1/2}\,a^+)\,,~\widehat{\sin\varphi}_D =
\frac{1}{2i}(a\,N^{-1/2}- N^{-1/2}\,a^+)\,,\ee or by Susskind and
Glogower, \begin{eqnarray} \widehat{\cos\varphi}_{SG}& =&
\frac{1}{2}[(N+1)^{-1/2}\,a+ a^+\,(N+1)^{-1/2}]\,, \\
\widehat{\sin\varphi}_{SG}& =& \frac{1}{2i}[(N+1)^{-1/2}\,a-
a^+\,(N+1)^{-1/2}]\,, \end{eqnarray} one sees immediately by which
kind of approximations of the expressions (85)  and (86) one
arrives at the disputable operators (87)-(89)! In order to do so
it is helpful to use the relations \be f(N)\,a^+ =
a^+f(N+1)\,,~a\,f(N)= f(N+1)\,a \ee for appropriate functions
$f(N)$ of the operator $N$, here applied to $f(N)=(N+k)^{-1}$
(provided $k>0 $). They are a consequence of the basic relations
(83). We then get for the expressions (85) and (86)
\begin{eqnarray}\widehat{\cos\varphi}&=& \frac{1}{2}[a^+F_k(N)+
F_k(N)a]\,, \\ \widehat{\sin\varphi} &=& \frac{i}{2}[a^+F_k(N)-
F_k(N)a]\,, \\ &&
F_k(N)=\frac{1}{2}\sqrt{N+2k}\left(\frac{1}{N+k}+\frac{1}{N+k+1}\right)\,.
\nonumber \end{eqnarray} As to the history of the operators
(87)-(89) see the reviews \cite{rev}.

From the expressions (91), (92) and the relation (90) one gets,
for instance, \be [\widehat{\cos\varphi},\widehat{\sin\varphi}]=
\frac{i}{2}\{[N+1]F_k^2(N)-N\,F_k^2(N-1)\} \ee which is the
operator version of the relation (34).

 It is interesting to have a
look at the expectation values of the operators $K_j$ and
$\widehat{\cos\varphi}$ and
 $\widehat{\sin\varphi}$ with respect to the conventional coherent
 states $|\alpha \rangle$ defined by \cite{squ} \be a\,|\alpha\rangle =
 \alpha\, |\alpha\rangle \,,~\alpha =r\,e^{i\beta}\, \in \,
 \Com\,,~~
 |\alpha \rangle =\sum_{n=0}^{\infty}
 \frac{\alpha^n}{\sqrt{n!}}e^{-|\alpha|^2/2}\,|n\rangle. \ee
 We get \begin{eqnarray} \langle \alpha|K_1|\alpha\rangle &=&
 r\cos\beta\, \langle \alpha|\sqrt{N+2k}|\alpha\rangle \,, \\
  \langle \alpha|K_2|\alpha\rangle
 &=&
- r\sin\beta\,\langle \alpha|\sqrt{N+2k}|\alpha\rangle\,, \\
\langle \alpha|K_3|\alpha\rangle &=& r^2 +k\,, \end{eqnarray}
where \be \langle \alpha|\sqrt{N+2k}|\alpha\rangle =
h_1^{(k)}(r^2) =
e^{-r^2}\sum_{n=0}^{\infty}\sqrt{n+2k}\,\frac{r^{2n}}{n!}\,. \ee
\\ From the expressions (91)-(92) one gets
\be \langle \alpha|\widehat{\cos\varphi}|\alpha\rangle =
\cos\beta\, h_2^{(k)}(r)\,,~~\langle
\alpha|\widehat{\sin\varphi}|\alpha\rangle = \sin\beta\,
h_2^{(k)}(r)\,, \ee where now \be
h_2^{(k)}(r)=\frac{r}{2}\,e^{-r^2}\,
\sum_{n=0}^{\infty}\sqrt{n+2k}\left(\frac{1}{n+k}+\frac{1}{n+k+1}
\right)\frac{r^{2n}}{n!}\,. \ee Due to the factor $\sqrt{n+2k}$
inside the sums (98) and (100) the functions
$h^{(k)}_j\,,\,j=1,2,$ do not seem to be summable in an elementary
way. Numerical inspections show that $h_2^{(k)} \leq 1$ for $k
\geq 0.5$ with $h_2^{(k)}(r) \to 1$ for $r \to \infty$ as it
should be. \\ Notice the differences between the relations (58)
and (50) on the one hand and the relations (95)-(97) on the other,
but also notice that - like in the case of the coherent states
$|k,z\rangle$ with the expectation values (64) and (65) - the
expectation values (99) are proportional to $\cos\beta$ and
$\sin\beta$ and the common function $h_2^{(k)}(r)$ which drops out
of the ratio  yielding $\cot\beta$ or $\tan\beta$, independent of
$k$ and $r$!

It is obvious that the realization (84) of the
$SO^{\uparrow}(1,2)$ Lie algebra does not represent an
interference situation we originally started from. It rather
represents the self-adjoint  Lie algebra generators in the state
space of the harmonic oscillator and provides self-adjoint
operators for $\sin$ and $\cos$. For $k=1/2$ the operator $K_3$
coincides with the Hamiltonian of the harmonic oscillator. \\
Thus,  the expressions (85) and (86) appear to present a whole
class of solutions for that old problem of quantizing the pair
``modulus'' and ``phase'' in a satisfactory manner for the
harmonic oscillator. The elements of the class are specified by
the parameter $k$ which characterizes an irreducible unitary
representation from the positive discrete series of the group
$SO^{\uparrow}(1,2)$ or of one of its covering groups!

 The expressions (84) form a highly non-linear realization of
the Lie algebra in terms of one pair of creation and annihilation
operators. In the literature it has been called a
``Holstein-Primakoff realization'' \cite{ge2}. These two were
among the first to express Lie algebra generators of $SU(2)$ in
terms of creation and annihilation operators \cite{ho}.

The following combination of $a$ and $a^+$ also fulfills the Lie
algebra (13) \cite{pe1,su11}: \be K_+ =\frac{1}{2}(a^+)^2\,,~K_-=
\frac{1}{2}a^2\,,~K_3=\frac{1}{2}(a^+a+1/2)\,. \ee As $K_-$
annihilates $|n=0\rangle$ as well as $|n=1\rangle$ the
representation decomposes into one with states of even quanta and
one with odd quanta. For even one has $k=1/4$ and for odd $k=3/4$.
Although the representation (101) is discussed quite frequently in
the quantum optics literature \cite{su11} it, too, does not appear
to be related to an interfering system as discussed above.

A realization of the Lie algebra (12) or (13) in terms of a pair
$a^+_j\,, a_j\,, j=1,2,$ of creation and annihilation operators
has been known for a long time \cite{bi,pe1,bo1} and has been
discussed frequently in quantum optics in the context of 2-mode
problems \cite{su11}, especially in connection with squeezed
states.

The operators \be K_3=\frac{1}{2}(a_1^+a_1+a_2^+a_2+1)~,~~
K_+=a_1^+a_2^+~, ~~K_-=a_1a_2 \ee obey the commutation relations
(13) and the tensor product $ \mbox{$\cal H$}^{osc}_1\otimes
\mbox{$\cal H$}^{osc}_2$ of the two harmonic oscillator Hilbert
spaces contains all the irreducible unitary representations of the
group $SU(1,1)$ (for which $k=1/2,1,3/2,\ldots$) in the following
way: Let $|n_i\rangle_j,~n_j \in \Nat_0,\, j=1,2,$ be the
eigenstates of the number operators $N_j$ generated by $a^+_j$
from the oscillator ground states. Then each of those two
subspaces of
 $\mbox{$\cal H$}^{osc}_1\otimes \mbox{$\cal H$}^{osc}_2= \{|n_1
 \rangle_1\otimes
 |n_2\rangle_2\}$ with fixed $|n_1-n_2|\neq 0$  contains an irreducible
  representation
 with $k=1/2+|n_1-n_2|/2=1,3/2,2,\ldots$ and for which the number
  $n$ in the eigenvalue $n+k$ is given by $n=\min\{n_1,n_2\}$. \\
   For the ``diagonal'' case
  $n_2=n_1$ one gets the unitary representation with $k=1/2$. \\
  Again, at first sight the operators (102) do not correspond to the
   interference
  (2) I started from, if one adheres to the correspondence $a_j \sim A_j\,,
   a^+ \sim \bar{A}_j$.
   This can be seen immediately: The operator
  $K_1=(a_1a_2+a^+_1a^+_2)/2$ corresponds to the classical
  quantity $(A_1A_2+\bar{A}_1\bar{A}_2)/2$, not to
  $(A_1\bar{A}_2+\bar{A}_1A_2)/2$, i.e.\ it corresponds to
  $\sqrt{I_1I_2}\cos(\varphi_1+\varphi_2)$ with the sum of the
  angle $\varphi_j$, not their difference! However, one should be
   careful here: if one of
   the angles, e.g.\ $\varphi_1$, vanishes the two situations are
   not so different.
    This situation corresponds to a certain fixed
   ``gauge'' of the angles, whereas the difference $\varphi=\varphi_2
   -\varphi_1$ is ``gauge invariant'':
   In addition, $K_3$ in
  Eq.\ (102) is given by the sum of the energies of the 2 modes,
  whereas $p$ in Eq.\ (6) is given by the square root of the
  product! Nevertheless the realization (102) has been used for other
  interfering devices \cite{kl} and our general analysis may be
  useful for those systems, too. It also might be of interest if
  $n_1=n_2$, i.e.\ if the energies of the 2 modes are the same as
  is the case classically if the two pinholes in Young's
  experiment have the same size, so that $I_2=I_1$.

  All these considerations do not impede our our original analysis of the
  observables in an interference pattern, e.g.\ in the case of the
  NFM set-up.  I repeat again that a group theoretical quantization
  like the one above
does $not$ suppose that there is a ``deeper'' conventional
canonical structure in terms of the usual $q$ and $p$ of $a$ and
$a^+$. It claims to provide an appropriate quantum framework for
topologically nontrivial symplectic manifolds like (11) by itself.
Quantum optics (or other quantum interference phenomena) may very
well be able to test such claims experimentally. In addition it
may test the identification (24) as an operator version of
$\cos\varphi$ and $\sin\varphi$. That definition is a new ansatz
{\em within} - not a basic ingredient {\em of} - the group
theoretical quantization scheme. \\ It even may turn out that the
operators $K_1$ and $K_2$ which correspond to the classical
observables $\sqrt{I_1I_2}\cos\varphi$ and
-$\sqrt{I_1I_2}\sin\varphi$ are actually more convenient to use
than the operators (24). The future will tell.
\section{Acknowledgements} The essential part of this work was
done while I was an invited guest of the CERN Theory Division. I
am very grateful for that invitation and I thank the Theory
Division and its head, Guido Altarelli, for their friendly and
supporting hospitality. I thank N.\ D\"{u}chting and S.\ Sint for
 numerical calculations. Very special thanks go to my
wife Dorothea!

\begin{thebibliography}{99}
\bibitem{rev} P.\ Carruthers and M.M.\ Nieto, Rev.\ Mod.\ Phys. {\bf
 40}, 411 (1968); \\
 Physica Scripta  {\bf  T48} (1993), edited by W.P.\ Schleich and
 S.M.\ Barnett; \\
 R.\ Lynch, Phys.\ Reports {\bf 256}, 367 (1995); \\
 M.\ Heni, M.\ Freyberger and W.P.\ Schleich, in {\em Coherence and
 Quantum Optics}
{\bf VII}, ed.\ by J.H.\ Eberly, L.\ Mandel and E.\ Wolf (Plenum
Press, New York and London, 1996) p.\ 239; \\ D.A.\ Dubin, M.A.\
Hennings and T.B.\ Smith, Intern.\ Journ.\ Mod.\ Phys.\ B {\bf 9},
2597 (1995); \\ D.T.\ Pegg and S.M.\ Barnett, Journ.\ Mod.\ Optics
{\bf 44}, 225 (1997); \\ D.-G.\ Welsch, W.\ Vogel and T.\
Opatrn\'{y}, Progr.\ in Optics {\bf 39}, 63 (1999)
\bibitem{Noh1} J.W.\ Noh, A.\ Foug\`{e}res and L.\ Mandel, Phys.\
Rev.\ Lett.\ {\bf 67}, 1426 (1991); Phys.\ Rev.\ A {\bf 45}, 424
(1992); Phys.\ Rev.\ A {\bf 46}, 2840 (1992); Phys.\ Rev.\ A {\bf
47}, 4535; 4541 (1993); Phys.\ Rev.\ Lett.\ {\bf 71}, 2579 (1993);
Phys.\ Rev.\ A {\bf 48}, 1719 (1993); Physica Scripta {\bf T48},
29 (1993)
\bibitem{Noh2} A.\ Foug\`{e}res, J.W.\ Noh, T.P.\ Grayson and L.\
Mandel, Phys.\ Rev.\ A {\bf 49}, 530 (1994)
\bibitem{Noh3}A.\ Foug\`{e}res, J.R.\ Torgerson and L.\ Mandel,
Optics Comm.\ {\bf 105}, 199 (1994)
\bibitem{Noh4}J.R.\ Torgerson and L.\ Mandel, Phys.\ Rev.\ Lett.\
{\bf 76}, 3939 (1995); Optics Comm.\ {\bf 133}, 153 (1997);
Physica Scripta {\bf T76}, 110 (1998)
\bibitem{is1}C.J.\ Isham  in {\em Relativity, Groups and Topology II}
 (Les Houches
Session XL), ed.\ by B.S.\ Dewitt and R.\ Stora (North-Holland,
Amsterdam etc., 1984) p.\ 1059; \\ V.\ Guillemin and S.\ Sternberg,
{\em Symplectic techniques in physics} (Cambridge University
Press, Cambridge etc., 1984)
\bibitem{bo1} M.\ Bojowald, H.A.\ Kastrup, F.\ Schramm and T.\ Strobl,
  Phys.\ Rev.\ D {\bf 62}, 044026 (2000),
gr-qc/9906105;\\ H.A.\ Kastrup, Ann.\ Physik (Leipzig) {\bf 9},
503 (2000), gr-qc/9906104
 \bibitem{lo} R.\ Loll, Phys.\ Rev.\ D {\bf 41}, 3785 (1990)
 \bibitem{ka1} H.A.\ Kastrup, quant-ph/0005033
 \bibitem{bo2} M.\ Bojowald and T.\ Strobl, J.\ Math.\ Phys.\ {\bf 41}, 2537
(2000), quant-ph/9908079; quant-ph/9912048;
\bibitem{ba1} A.O.\ Barut and L.\ Girardello, Commun.\ math.\ Phys.\
{\bf 21}, 41 (1971)
\bibitem{lou} W.H.\ Louisell, Phys.\ Lett.\ {\bf 7}, 60 (1963)
\bibitem{ge1} C.C.\ Gerry, Phys.\ Rev.\ A {\bf 38}, 1734 (1988)
\bibitem{ag1} G.S.\ Agarwal, J.\ Opt.\ Soc.\ Am.\ B {\bf 5}, 1940
(1988)
\bibitem{vo1} A.\ Vourdas, Phys.\ Rev.\ A {\bf 41}, 1653 (1990)
\bibitem{bri} C.\ Brif, Quantum Semiclass.\ Opt.\ {\bf 7}, 803
(1995)
\bibitem{kl1} J.R.\ Klauder and B.-S.\ Skagerstam, {\em Coherent States --
Applications in Physics and Mathematical Physics}, (World
Scientific Publ.\ Co., Singapore, 1985)
\bibitem{pe1} A.\ Perelomov, {\em Generalized Coherent States and Their
Applications} (Sprin\-ger-Verlag, Berlin etc., 1986)
\bibitem{zha} W.-M.\ Zhang,
D.H.\ Feng and R.\ Gilmore, Rev.\ Mod.\ Phys.\ {\bf 62}, 867
(1990)
\bibitem{ha1} B.C.\ Hall, Contemp.\ Mathem.\ {\bf 260}, 1 (2000),
quant-ph/9912054
\bibitem{tri} D.A.\ Trifonov, J.\ Math.\ Physics {\bf 35}, 2297
(1994);\\
 C.\ Brif, Intern.\ J.\ of Theor.\ Physics {\bf 36}, 1651 (1997);
 \\
 D.A.\ Trifonov, JOSA A {\bf 17} (No.\ 12), 2486 (2000), quant-ph/0012072
\bibitem{su11} A selection out of many papers is: \\
 K.\ W\'{o}dkiewicz and J.H.\ Eberly, Journ.\ Opt.\ Soc.\ Am.\ B {\bf
 2}, 458 (1985); \\ B.L.\ Schumaker and C.M.\ Caves, Phys.\ Rev.\ A {\bf
31}, 3093 (1985);\\ R.F.\ Bishop and A.\ Vourdas, Journ.\ Phys.\ A:
Math.\ Gen.\ {\bf 19}, 2525 (1986) and {\bf 20}, 3727 (1987);\\ B.\
Yurke, S.L.\ McCall and J.R.\ Klauder, Phys.\ Rev.\ A {\bf 33},
4033 (1986); \\  J.\ Katriel, A.I.\ Solomon, G.\ D'Ariano and M.\
Rasetti, Phys.\ Rev.\ D {\bf 34}, 2332 (1986); \\
 C.C.\ Gerry, Phys.\ Rev.\ A {\bf 35}, 2146 (1987);\\ G.S.\ Agarwal,
  Ref.\ \cite{ag1}; \\ M.\
Hillery, Phys.\ Rev.\ A {\bf 40}, 3147 (1989); \\ C.C.\ Gerry,
Journ.\ Opt.\ Soc.\ Am.\ B {\bf 8}, 685 (1990); \\ J.A. Bergou, M.\
Hillery and D.\ Yu, Phys.\ Rev.\ A {\bf 43}, 515 (1991); \\ A.\
Vourdas, Phys.\ Rev.\ A {\bf 46}, 442 (1992); \\ H.-Y.\ Fan and X.\
Ye, Phys.\ Lett.\ A {\bf 175}, 387 (1993);\\ M.\ Ban, Phys.\ Rev.\ A
{\bf 47}, 5093 (1993); \\ M.M.\ Nieto and D.R.\ Truax, Phys.\ Rev.\
Lett.\ {\bf 71}, 2843 (1993); \\
 U.\ Leonhardt, Phys.\ Rev.\ A {\bf 49}, 1231 (1994); \\ G.S.\
Prakash and G.S.\ Agarwal, Phys.\ Rev.\ A {\bf 50}, 4258 (1994);
\\ B.A.\ Bambah and G.S.\ Agarwal, Phys.\ Rev.\ A {\bf 51}, 4918
(1995); \\ C.C.\ Gerry and R.\ Grobe, Phys.\ Rev.\ A {\bf 51},
1698 and 4123 (1995); \\ C.\ Brif, Ann.\ Phys.\ (N.Y.) {\bf 251},
180 (1996); \\ H.-C.\ Fu and R.\ Sasaki, Phys.\ Rev.\ A {\bf 53},
3836 (1996);
\\ S.-C.\ Gou, J.\ Steinbach and P.L.\ Knight, Phys.\ Rev.\ A {\bf
54}, 4315 (1996);
\\ C.\ Brif, Ref.\ \cite{bri};\\ X.-G.\ Wang, Intern.\ J.\ Mod.\
Physics B {\bf 14}, 1093 (2000); Opt.\ Comm.\ {\bf 178}, 365
(2000); J.\ Opt.\ B: Quantum Semiclass. Opt.\ {\bf 2}, 534 (2000);
\\ X.-G.\ Wang, B.C.\ Sanders and S.-H.\ Pan, J.\ Phys.\ A: Math.\
Gen. {\bf 33}, 7451 (2000)
\bibitem{squ} As to the vast literature on squeezed states see
Refs.\ \cite{su11,zha,bri} and  text books on quantum optics,
e.g.\ \\ D.F.\ Walls and G.J.\ Milburn, {\em Quantum Optics}
(Springer-Verlag, Heidelberg etc., 1994); \\ L.\ Mandel and E.\
Wolf, {\em Optical Coherence and Quantum Optics} (Cambridge
University Press, Cambridge etc., 1995); \\ U.\ Leonhardt, {\em
Measuring the Quantum State of Light} (Cambridge University Press,
Cambridge etc., 1997); \\ V.\ Pe\v{r}inov\'{a}, A.\ Luk\v{s} and
J.\ Pe\v{r}ina, {\em Phase in Optics} (World Scientific Publ.\
Co., Singapore, 1998); \\ W.\ Vogel,
 D.-G.\
Welsch and S.\ Wallentowitz, {\em Quantum Optics. An
Introduction}, 2nd Ed.\ (Wiley-VHC Verlag, Weinheim, 2001)
\bibitem{er1} A.\ Erd\'{e}lyi et al.\ (Eds.), {\em Higher Transcendental
 Functions
  II} (McGraw-Hill Book Co.\, Inc., New York etc., 1953) ch.\ VII
\bibitem{GR} Ref.\ \cite{er1}, p.\ 51, Eq.\ (27)
\bibitem{lu1} Y.L.\ Luke, {\em Integrals of Bessel Functions} (McGraw-Hill
 Book Co.,
New York etc.\ 1962) p.\ 85: 3.9., formula (2)
\bibitem{lu2} Ref.\ \cite{lu1}, p.\ 55: 2.5., formula (10)
\bibitem{ree} M.\ Reed and B.\ Simon, {\em Methods of Modern Mathematical
 Physics,
I: Functional Analysis} (Academic Press, New York and London, 1972)
p.\ 192: Theorem VI.6 and p.\ 216: problem 9
\bibitem{you} See, e.g.\ M.\ Born and E.\ Wolf, {\em Principles of
Optics}, 7th Ed.\ (Cambridge University Press, Cambridge, 1999);
\\
D.F.\ Walls and G.J, Milburn, {\em Quantum Optics} (Springer-Verlag,
 Heidelberg etc.\ 1994)
\bibitem{wa} A nice discussion of the quantum optics of Young's
experiment gives D.F.\ Walls, Amer. J.\ of Physics {\bf 45}, 952
(1977); as to a very recent associated experiment see \\ A.F.\
Abouraddy, M.B.\ Nasr, B.E.A.\ Saleh, A.V.\ Sergienko and M.C.\
Teich, Phys.\ Rev.\ A {\bf 63}, 063803 (2001)
\bibitem{go} P.\ Goddard and D.\ Olive, Intern.\ J.\ of Mod.\
Physics A {\bf 1}, 303 (1986)
\bibitem{ge2} The realization (84) was apparently first discussed
by L.D.\ Mlodinow and N.\ Papanicoulaou, Ann.\ Phys.\ (N.Y.) {\bf
128}, 314 (1980); \\ see also \\ C.C.\ Gerry, J.\ Phys.\ A: Math.\
Gen.\ {\bf 16}, L1 (1983); \\ J.\ Katriel, A.I.\ Solomon, G.\
D'Ariano and M.\ Rasetti, Phys.\ Rev.\ D {\bf 34}, 2332 (1986);
\\ C.C.\ Gerry and R.\ Grobe, Quantum Semiclass.\ Opt.\ {\bf 9},
59 (1997);
\\ A.\ W\"{u}nsche, Acta physica slovaca {\bf 49}, 771 (1999)
\bibitem{ho} T.\ Holstein and H.\ Primakoff, Phys.\ Rev.\ {\bf
58}, 1098 (1940)
\bibitem{bi}S.\ Goshen (Goldstein) and H.J.\ Lipkin, Ann.\ Phys.\
(N.Y.) {\bf 6}, 301 (1959); \\  H.J.\ Lipkin, {\em Lie Groups for
Pedestrians} (North-Holland Publ.\ Co., Amsterdam, 1965) ch.\ 5;
\\ W.J.\ Holman, III and L.C.\ Biedenharn, Jr., Ann.\ Phys.\
(N.Y.) {\bf 39}, 1 (1966); \\ B.G.\ Wybourne, {\em Classical
Groups for Physicists} (John Wiley \& Sons, New York etc., 1974)
ch.\ 17
\bibitem{kl} B.\ Yurke, S.L.\ McCall and J.R.\ Klauder, Phys.\ Rev.\
 A {\bf 33},
4033 (1986); \\ U.\ Leonhardt, Phys.\ Rev.\ A {\bf 49}, 1231
(1994); \\ C.\ Brif and A.\ Mann, Phys.\ Rev.\ A {\bf 54}, 4505
(1996); \\ V.\ Pe\v{r}inov\'{a}, A.\ Luk\v{s} and J.\
K\v{r}epelka, J.\ Opt.\ B: Quantum Semiclass.\ Opt.\ {\bf 2}, 81
(2000)
\end{thebibliography}
  \end{document}